\documentstyle[12pt]{article}
\begin{document}
\topmargin -0.2cm \oddsidemargin -0.2cm \evensidemargin -1cm
\textheight 22cm \textwidth 12cm

\title{ Superfluid Component of Electromagnetic Field  and New Fundamental Constant in The Nature.}  
\author{ V.N.Minasyan \\
Yerevan, Armenia}

\date{\today}

\maketitle

\begin{abstract}  
First, the description of a quantization local electromagnetic field is proposed 
by presentation of quantum form of Maxwell equations in the vacuum which describes the electromagnetic field by the model of a Bose-gas consisting of the interacting fundamental neutral Bose-particles (light bosons) with spin one and a finite mass $m$. The later represents as a fundamental constant $ m=\frac{m_e e^4}{2\hbar^2 c^2}=2.4 \cdot 10^{-35} kg $. The light bosons of electromagnetic field induce the Bose-excitations, which are the photon modes. In this letter, we show that the Bose gas of light bosons undergoes a phase transition at low temperatures to condition in which the zero-momentum quantum state is occupied by a finite fraction of the light bosons. This momentum-condensed phase represents as a superfluid component for electromagnetic field. The later is absent at the transition temperature, which is found in this letter.
\end{abstract}


\vspace{100mm} 

\vspace{5mm}

{\bf 1. INTRODUCTION.} 

\vspace{5mm}
The motivation for our theoretical study of the quantization scheme for local electromagnetic field in the vacuum is an attempt at a microscopic understanding of properties of the electromagnetic field. As originally known, at investigation of spectrum energy of 'Black Body' by the classic Maxwell equations, it was an appearance so-called  "ultraviolet catastrophe". To remove a later, the Plank proposes to consider the electromagnetic filed by the model of ideal Bose gas consisting of massless photons with spin one. 

In this context, the Dirac [1] proceed an investigation of the important problem as an obtaining of Plank photon-gas by way of introducing quantization scheme for local electromagnetic field. In this respect, Dirac proposed the theoretical description of quantization local electromagnetic field in the vacuum within of the model of Bose-gas consisting of a local plane electromagnetic waves, which are propagated by speed  $c$ in vacuum. In this context, the Dirac introduces a quantization scheme by considering a vector of electromagnetic field in space of wave-vector as the Bose operators of  "creation" and "annihilation" of plane waves with spin one.     

In this letter, we prove that the quantized classic equations of Maxwell, presented in vacuum, are not able to describe a quantization scheme for electromagnetic field within presentation the Plank photon-gas because besides introducing quantization scheme, there is a necessary to include also new solution of Maxwell equations. Thus, we propose a presentation of new quantized form of Maxwell equations, which allow obtaining of the Plank photon-gas.   

For beginning, we present the Maxwell equations in the zero-vacuum:

\begin{equation}
curl {\vec {H}} -\frac{1}{c}\frac{d {\vec{E}}}{d t}=0
\end{equation}

\begin{equation}
curl {\vec {E}} +\frac{1}{c}\frac{d {\vec{H}}}{d t}=0
\end{equation}

\begin{equation}
div {\vec {E}} =0
\end{equation}

\begin{equation}
div {\vec {H}} =0
\end{equation}

where  $\vec {E}=\vec {E}(\vec {r},t)$ and $\vec {H}=\vec {H}(\vec {r},t)$ are, respectively, the local electric and magnetic fields presented in dependence of the coordinate $\vec {r}$ and current time $t$; $c$ is the velocity of wave in vacuum. 

The Hamiltonian of radiation $\hat{H}_R$ is determined as: 

\begin{equation} 
\hat{H}_R =\frac{1}{8\pi V}\int \biggl( E^2+H^2\biggl) dV 
\end{equation}

\vspace{5mm}

{\bf 11. QUANTIZATION SCHEME FOR MAXWELL QUATIONS.} 

\vspace{5mm}

Following to the Dirac theory, who first introduced the quantization shame for electromagnetic fields we use of the wave-equations for $\vec {E}$:

$$ 
\nabla^2 {\vec {E} }-\frac{1}{c^2}\frac{d^2 \vec{E}}{d t^2}=0 
$$

The given equation has a following solution:

\begin{equation} 
\vec {E}= \frac{1}{V}\sum_{\vec{k}}\biggl(
\vec { E } _{\vec{k}} e^{i(
\vec{k}\vec{r} + kc t )} +\vec { E }^{+}_{\vec{k}}
e^{-i(\vec{k}\vec{r} + kc t)}\biggl)
\end{equation}

where  $\vec { E } ^{+}_{\vec{k}}$  and $\vec { E } _{\vec{k}}$ are, respectively, the Fourier components of vectors electric plane local field in space of wave vector $\vec{k}$, which are determined by the vector Bose-operators "creation" and "annihilation" the Bose-waves with spin one and energy $\hbar k c$, These plane waves are propagated by speed $c$ in forward by direction of wave vector $\vec{k}$. The vector Bose-operators of the Bose-waves satisfy to the Bose-commutation relations:

$$
\biggl[\vec { E }_{\vec{k}},\vec { E }^{+}_{\vec{k}^{'} }\biggl]_{-} =
\delta_{\vec{k},\vec{k^{'}}} 
$$

$$
\biggl[\vec { E }_{\vec{k}}, \vec { E }_{\vec{k^{'}}}\biggl]_{-}= 0
$$

$$
\biggl[\vec { E }^{+}_{\vec{k} }, \vec { E }^{+}_{\vec{k^{'}}}\biggl]_{-}= 0
$$

Consequently, by inserting of a values (6)  into (5), by using of 

\begin{equation}
\frac{1}{V}\int e^{i\vec{k}\vec{r}} dV=\delta_{\vec{k}}
\end{equation}

we obtain the Hamiltonian radiation $\hat{H}_R $ in a following way by using of the condition $E^2=H^2$:

\begin{eqnarray} 
\hat{H}_R &=&
\frac{1}{4\pi V }
\int E^2 dV =
\nonumber\\
&=&\frac{1}{4\pi }\sum_{\vec{k},\vec{k}_1} \delta (k+ {k}_1) \biggl(
\vec {E} _{\vec{k}} + \vec  {E}^{+}_{-\vec{k}}\biggl) \biggl(\vec {E}_{\vec{k}_1}  +\vec  {E}^{+}_{-\vec{k}_1}\biggl) =
\nonumber\\
&=&\frac{1}{4\pi }\sum_{\vec{k}} \biggl(\vec {E} _{\vec{k}}  +\vec  {E}^{+}_{-\vec{k}}\biggl)
\biggl(\vec {E} _{-\vec{k}} +\vec  {E}^{+}_{\vec{k}}\biggl)
\end{eqnarray}

Thus, 
\begin{equation} 
\hat{H}_R =\frac{1}{4\pi }\sum_{\vec{k}} \biggl(2\vec  {E}^{+}_{\vec{k}}\vec {E} _{\vec{k}}  +\vec  {E}^{+}_{\vec{k}}\vec {E}^{+} _{-\vec{k}} +\vec  {E}_{-\vec{k}}\vec {E}_{\vec{k}} +1\biggl)
\end{equation}

The evaluation of energy levels of the operators $\hat{H}_R$ in Eq. (9)
within diagonal form, we apply a new linear transformation for vector Bose-operators which is a similar to the Bogoliubov linear transformation of scalar Bose-operators [2]: 
 
\begin{equation} 
\vec {E}_{\vec{k}}=\frac{\vec {e}_{\vec{k}} + 
L_{\vec{k}}\vec {e}^{+}_{-\vec{k}}} {\sqrt{1-L^2_{\vec{k}}}}
\end{equation}

where $L_{\vec{k}}$ is the real symmetrical functions  
of  a wave vector $\vec{k}$. 
 
Thus, the diagonal form of operator Hamiltonian $\hat{H}_R$ takes a zero value:  
\begin{equation}
\hat{H}_R= 2\sum_{\vec{k}}\xi_{\vec{k}}\vec {e}^{+}_{\vec{k}}\vec {e}_{\vec{k}}+\frac{1}{4\pi }\sum_{\vec{k}}1
\end{equation}

where we infer that the Bose-operators 
$\vec {e}^{+}_{\vec{k}}$ and  $\vec {e}_{\vec{k}}$ are, respectively, 
the "creation" and "annihilation" vector operators of 
free quasi-bosons of electric field with energies 
$\xi_{\vec{k}}=\sqrt{\frac{1}{16\pi^2 }-\frac{1}{16\pi^2 }}=0$. Therefore,

$$
\hat{H}_R= \frac{1}{4\pi }\sum_{\vec{k}}1
$$
which is a constant. This reasoning implies that the introduction of quantisation scheme for the classic Maxwell equations cannot describe a microscopic property of electromagnetic field.

\vspace{5mm}

{\bf 111.  THEORY OF DIRAC}.
\vspace{5mm}

In this respect, the Dirac proposed to examine a quantization scheme of electromagnetic field by introducing of the vector potential for local electromagnetic field $\vec {A}(\vec{r},t)$:

\begin{equation} 
\vec {H}= curl  {\vec {A}}
\end{equation} 
and
\begin{equation} 
\vec {E}=- \frac{1}{c}\frac{d {\vec{A}}}{d t}
\end{equation} 
  
which by inserting in (1)-(4), determines a wave-equation:

\begin{equation} 
\nabla^2 {\vec {A} }-\frac{1}{c^2}\frac{d^2 \vec{A}}{d t^2}=0 
\end{equation} 
with condition
\begin{equation} 
div {\vec {A} }=0
\end{equation} 
The solution of (14) is presented by a following form:

\begin{eqnarray} 
\vec {A}(\vec{r},t)& =&\int  \biggl(\vec {A} _{\vec{k}} \exp^{i(
\vec{k}\vec{r} + kc t )} +\vec {A}^{+}_{\vec{k}}
\exp^{-i(\vec{k}\vec{r} + kc t)} \biggl) d^3 k=\nonumber\\
&=&\sum_{\vec{k}}\biggl(
\vec {A} _{\vec{k}} \exp^{i(
\vec{k}\vec{r} + kc t )} +\vec {A}^{+}_{\vec{k}}
\exp^{-i(\vec{k}\vec{r} + kc t)}\biggl)
\end{eqnarray} 

where $\vec {A} ^{+}_{\vec{k}}$  and $\vec {A} _{\vec{k}}$ are, respectively, the Fourier components of vector potentials electromagnetic field which are considered as the vector Bose-operators "creation" and "annihilation" of a Bose-plane wave with spin one. 

Obviously, we have an expression for component $ {H}_x$ on coordinate $x$:

\begin{eqnarray}
{H}_x& =&\biggl (curl {\vec {A}}\biggl )_x= \frac{d A_z}{dy}-\frac{d A_y}{dz}=
\nonumber\\
&=&\biggl (i\sum_{\vec{k}}\vec {k}\times \biggl(
\vec {A} _{\vec{k}} \exp^{i(
\vec{k}\vec{r} + kc t )} -\vec {A}^{+}_{\vec{k}}
\exp^{-i(\vec{k}\vec{r} + kc t)}\biggl) \biggl)_x
\end{eqnarray}
Then,

\begin{equation}
\vec {H}=curl {\vec {A}}=i\sum_{\vec{k}}\vec {k}\times \biggl(
\vec {A} _{\vec{k}} \exp^{i(
\vec{k}\vec{r} + kc t )} -\vec {A}^{+}_{\vec{k}}
\exp^{-i(\vec{k}\vec{r} + kc t)}\biggl)
\end{equation}

To find $H^2$, we use of a supporting formulae from textbook [3] 
$$
\biggl[\vec {a} \times \vec {b}\biggl]\cdot \biggl[\vec {c} \times \vec {d}\biggl]=( \vec {a}\cdot \vec {c})(\vec {b}\cdot \vec {d}) -
(\vec {a}\cdot \vec {d})(\vec {b}\cdot \vec {c})
$$

which leads to following form for $\frac{1}{8\pi V}\int H^2dV $
at application (16) and (7), with using of the condition of transverse wave $\vec{k}\cdot \vec {A}_{\vec{k}}=0$, we posses 
\begin{eqnarray}
\frac{1}{8\pi V}\int H^2dV&=&-\frac{1}{8\pi}\sum_{\vec{k},\vec{k}_1}\delta_{\vec{k} + \vec{k}_1}  
\vec {k}\vec  {k}_1 
\biggl(\vec {A}_{\vec{k}}  -\vec {A}^{+}_{-\vec{k}}\biggl) 
\biggl(\vec {A}_{\vec{k}_1}  -\vec  {A}^{+}_{-\vec{k}_1}\biggl)=
\nonumber\\
&=&  \frac{1}{8\pi }\sum_{\vec{k}} k^2\biggl(\vec {A} _{\vec{k}}  -\vec  {A}^{+}_{-\vec{k}}\biggl)
\biggl(\vec {A} _{-\vec{k}}  -\vec  {A}^{+}_{\vec{k}}\biggl)
\end{eqnarray}

We now calculate the part of the Hamiltonian radiation in (5) 
\begin{equation}
\frac{1}{8\pi V}\int  E^2 dV=\frac{1}{8\pi c^2 V }
\int \biggl(\frac{d {\vec{A}}}{d t}\biggl)^2 dV 
\end{equation}  

At calculation of value $\frac{d {\vec{A}}}{d t}$, we use of a suggestion proposed by Dirac which implies a consideration of current time $t=0$  [1]:

\begin{equation} 
\frac{d {\vec{A}}}{d t}=ic\sum_{\vec{k}}k\biggl(
\vec {A} _{\vec{k}} -\vec {A}^{+}_{\vec{k}}\biggl)
e^{i\vec{k}\vec{r}}
\end{equation}

Inserting value of $\frac{d {\vec{A}}}{d t}$ from Eq.(21) into Eq.(20), we find the Hamiltonian of radiation $\hat{H}_R $ by following form:

\begin{eqnarray} 
\frac{1}{8\pi c^2 V }
\int \biggl(\frac{d {\vec{A}}}{d t}\biggl)^2 dV& =&
-\frac{ 1}{8\pi }\sum_{\vec{k},\vec{k}_1} \delta _{k+ {k}_1} |k| \cdot |k_1|  \biggl(
\vec {A} _{\vec{k}} - \vec  {A}^{+}_{-\vec{k}}\biggl) \biggl(\vec {A}  _{\vec{k}_1}  -\vec  {A}^{+}_{-\vec{k}_1}\biggl) =
\nonumber\\
&=& - \frac{1}{8\pi }\sum_{\vec{k}} k^2\biggl(\vec {A} _{\vec{k}}  -\vec  {A}^{+}_{-\vec{k}}\biggl)
\biggl(\vec {A} _{-\vec{k}}  -\vec  {A}^{+}_{\vec{k}}\biggl)
\end{eqnarray}
Thus, by using of results (19) and (22), we obtain the Dirac Hamiltonian

\begin{equation} 
\hat{H}_R =\frac{1}{8\pi V}\int  E^2 dV +\frac{1}{8\pi V}\int H^2dV =0 
\end{equation} 

which is not able to describe the Plank photon gas. This fact allows us to suggest that it needs to find a new solution of Maxwell equations, which could provide the description of the Plank photon gas.  
 
\vspace{5mm}
{\bf 1V. FRESH THEORY}
\vspace{5mm}

To solve a problem connected with a quantization electromagnetic field, we propose the quantized equations of Maxwell.  For beginning, we search the solution of (1)-(4) by following way:  

\begin{equation} 
\vec {E}=- \frac{\alpha}{c}\cdot
\frac{d {\vec{H}_0}}{d t}+\beta\cdot \vec {E}_0
\end{equation} 
and
\begin{equation} 
\vec {H}= \alpha \cdot  curl{
\vec {H}_0} + \beta \vec{H}_0
\end{equation} 

second quantization wave functions of "creation" and "annihilation" of free Bose- particles for one boson in the space of coordinate $\vec{r}$

where $\alpha $ and $\beta $ are the constants which we obtain in the bellow by using of a physical property of electromagnetic field; $\vec {E}_0=\vec {E}_0(\vec {r},t)$ and $\vec {H}_0=\vec {H}_0 (\vec {r},t)$ are, respectively, determined as the vectors second quantization wave functions for one Bose particle of electromagnetic field with spin one and  mass $m$. In this context, we claim that the vectors of local electric $\vec {E}_0$ and magnetic $\vec {H}_0$ fields, presented by equations (24) and (25), satisfy to the equations of Maxwell in vacuum which here describe the states of the Bose particles:

\begin{equation}
curl {\vec {H}_0} -\frac{1}{c}\frac{d {\vec{E}_0}}{d t}=0
\end{equation}

\begin{equation}
curl {\vec {E}_0} +\frac{1}{c}\frac{d {\vec{H}_0}}{d t}=0
\end{equation}

\begin{equation}
div {\vec {E}_0} =0
\end{equation}

\begin{equation}
div {\vec {H}_0} =0
\end{equation}
In this context, by using of (26), we can rewrite (23) as 
\begin{equation} 
\vec {H}= \frac{\alpha }{c}\frac{d {\vec{E}_0}}{d t} + \beta \vec{H}_0
\end{equation}

By presentation of new terms $E_0$ and $H_0$, the Hamiltonian of radiation $\hat{H}_R$ in (5) takes a following form: 

\begin{eqnarray} 
\hat{H}_R &=&\frac{1}{8\pi V}\int \biggl( E^2+H^2\biggl) dV = \frac{1}{8\pi V}\int \biggl [\biggl(- \frac{\alpha }{c}\frac{d {\vec{H}_0}}{d t}+\nonumber\\
&+&\beta \vec {E}_0\biggl)^2+\biggl (\frac{\alpha }{c}\frac{d {\vec{E}_0}}{d t}+ \beta \vec {H}_0\biggl)^2\biggl] dV =
\hat{H}_e+\hat{H}_h
\end{eqnarray}

where

\begin{equation} 
\hat{H}_e=\frac{1}{8\pi V}\int \biggl [\biggl(\frac{\alpha}{c}
\frac{d {\vec{E}_0}}{d t}\biggl)^2+
\beta^2\vec {E}^2_0\biggl]dV
\end{equation}

\begin{equation} 
\hat{H}_h=\frac{1}{8\pi V}\int \biggl [\biggl(\frac{\alpha}{c}
\frac{d {\vec{H}_0}}{d t}\biggl)^2+
\beta^2\vec {H}^2_0\biggl]dV
\end{equation}

Obviously, the equations (26)-(29) lead to a following wave-equation:

\begin{equation} 
\nabla^2 {\vec {E}_0}-\frac{1}{c^2}\frac{d^2 \vec{E}_0}{d t^2}=0 
\end{equation}
and
\begin{equation} 
\nabla^2 {\vec {H}_0}-\frac{1}{c^2}\frac{d^2 \vec{H}_0}{d t^2}=0 
\end{equation}
which in turn have following solutions:

\begin{equation} 
\vec {E}_0= \frac{1}{V}\sum_{\vec{k}}\biggl(
\vec {E} _{\vec{k}} e^{i(
\vec{k}\vec{r} + kc t )} +\vec {E}^{+}_{\vec{k}}
e^{-i(\vec{k}\vec{r} + kc t)}\biggl)
\end{equation}

\begin{equation} 
\vec {H}_0= \frac{1}{V}\sum_{\vec{k}}\biggl(
\vec {H} _{\vec{k}} e^{i(
\vec{k}\vec{r} + kc t )} +\vec {H}^{+}_{\vec{k}}
e^{-i(\vec{k}\vec{r} + kc t)}\biggl)
\end{equation} 

where  $\vec { E } ^{+}_{\vec{k}}$, $\vec { H } ^{+}_{\vec{k}}$  and  $\vec {E} _{\vec{k}}$, $\vec {H} _{\vec{k}}$ are, respectively, the second quantzation vectors wave functions, which are represented as the vector Bose-operators "creation" and "annihilation" of the Bose-particles of electric and magnetic waves with spin one. 

We now insert a value of $\vec {E}_0$ from (36) into (33), and then: 

\begin{eqnarray} 
\frac{1}{8\pi V}\int \biggl [\biggl(\frac{\alpha}{c}
\frac{d {\vec{E}_0}}{d t}\biggl)^2dV&=&-\frac{\alpha^2}{8\pi}\sum_{\vec{k},\vec{k}_1} 
\delta_{\vec{k}+\vec{k}_1} |k| \cdot |k_1|  \biggl(
\vec {E}_{\vec{k}} - \vec{E}^{+}_{-\vec{k}}\biggl) 
\biggl(\vec{E}_{\vec{k}_1} -\vec{E}^{+}_{-\vec{k}_1}\biggl) =
\nonumber\\
&=&-\frac{\alpha^2}{8\pi}\sum_{\vec{k}}k^2
\biggl(\vec{E}_{\vec{k}} -\vec{E}^{+}_{-\vec{k}}\biggl)
\biggl(\vec{E}_{-\vec{k}} -\vec{E}^{+}_{\vec{k}}\biggl)
\end{eqnarray}

Consequently, within introducing of assumption that the 
term with square wave vector $k^2$ describes the kinetic 
energy of free Bose-particles of electromagnetic field with mass $m$ by definition $\frac{\alpha^2 k^2}{4\pi}=\frac{\hbar^2 k^2}{2m}$,  we find a constant $\alpha = \frac{\hbar \sqrt{2\pi}}{\sqrt{m}}$. Then, we posses:

\begin{eqnarray} 
\frac{1}{8\pi V}\int \biggl [\biggl(\frac{\alpha}{c}
\frac{d {\vec{E}_0}}{d t}\biggl)^2dV &=&\sum_{\vec{k}}\frac{\hbar^2k^2}{2m}\vec {E}^{+}_{\vec{k}}\vec {E}_{\vec{k}}- 
\sum_{\vec{k}}\frac{\hbar^2k^2}{4m}
\biggl (\vec {E}^{+}_{\vec{k}}
\vec {E}^{+}_{-\vec{k}}+   
\vec {E}_{-\vec{k}}\vec {E}_{\vec{k}}\biggl)- \nonumber\\
&-&\sum_{\vec{k}}\frac{\hbar^2k^2}{4m}
\end{eqnarray}

As we see the first term in right pat of (39) represents as the kinetic energy of the Bose gas consisting of the Bose-particles of electromagnetic field but the second term in right pat of (39) describes the term of the interaction between particles. 
In this context,  the part $\hat{H}_e$ in (32) takes a following form:
\begin{eqnarray} 
\frac{1}{8\pi V}\int 
\beta^2\vec {E}^2_0dV &=&
\frac{\beta^2}{8\pi}\sum_{\vec{k},\vec{k}_1}
\delta(\vec{k}+ \vec{k}_1)\biggl(\vec {E} _{\vec{k}} +
\vec  {E}^{+}_{-\vec{k}}\biggl) 
\biggl(\vec {E}_{\vec{k}_1} +\vec  {E}^{+}_{-\vec{k}_1}\biggl) 
=\nonumber\\
&=&\frac{\beta^2}{8\pi}\sum_{\vec{k}} 
\biggl(\vec {E} _{\vec{k}}+\vec  {E}^{+}_{-\vec{k}}\biggl)
\biggl(\vec {E} _{-\vec{k}} +\vec{E}^{+}_{\vec{k}}\biggl)=
\nonumber\\
&=&\frac{\beta^2}{8\pi}\sum_{\vec{k}} 
\biggl(2\vec  {E}^{+}_{\vec{k}}\vec {E} _{\vec{k}}+
\vec {E} _{\vec{k}}\vec {E} _{-\vec{k}} +\vec{E}^{+}_{-\vec{k}}\vec{E}^{+}_{\vec{k}}\biggl)+ \frac{\beta^2}{8\pi}\sum_{\vec{k}}1
\end{eqnarray}
Consequently, the  operator $\hat{H}_e$ is presented by a following form:

\begin{equation} 
\hat{H}_e=
\sum_{\vec{k}}\biggl (\frac{\hbar^2
k^2}{2m }+
\frac{\beta^2}{4\pi }\biggl )
\vec{E}^{+}_{\vec{k}}\vec{E}_{\vec{k}}+ 
\frac{1}{2V}\sum_{\vec{k}}\hat{U}_{\vec{k}}
\biggl (\vec{E}^{+}_{\vec{k}}
\vec{E}^{+}_{-\vec{k}}+   
\vec{E}_{-\vec{k}}\vec{E}_{\vec{k}}\biggl)
\end{equation}

where $\hat{U}_{\vec{k}}=-\frac{\hbar^2k^2 V}{2m }+
\frac{\beta^2V}{4\pi }$ in the second term in right side of (41) describes the interaction between the Bose-particles. We claim that the inter-particle interaction $\hat{U}_{\vec{k}}$ represents as a repulsive $\hat{U}_{\vec{k}}>0$ in the space of wave vector $\vec{k}$. This assumption leads to the condition for wave numbers  $k\leq k_0=\frac{\beta }{\hbar}\sqrt{\frac{m}{2\pi}}$ where $k_0$ is the boundary maximal wave number which provides that the existence of the interaction energy $\hat{U}_{\vec{r}}$ between two light bosons in the coordinate space (the form of  $\hat{U}_{\vec{r}}$ will be presented in section V): 

\begin{equation}
\hat{U}_{\vec{r}}=\frac{1}{V}\sum_{\vec{k}} \hat{U}_{\vec{k}}\cdot e^{i\vec{k}\vec{r}}
\end{equation}  

Obviously, the sum in (42) diverges, within introducing the concept of a boundary wave number $k_0$ for electromagnetic field. The light bosons with wave number exceeding the boundary wave number $k_0$ do not exist. 

In analogy manner, we can find

\begin{equation} 
\hat{H}_h=
\sum_{\vec{k}}\biggl (\frac{\hbar^2
k^2}{2m }+
\frac{\beta^2}{4\pi }\biggl )
\vec{H}^{+}_{\vec{k}}\vec{H}_{\vec{k}}+ 
\frac{1}{2V}\sum_{\vec{k}}\hat{U}_{\vec{k}}
\biggl (\vec{H}^{+}_{\vec{k}}
\vec{H}^{+}_{-\vec{k}}+   
\vec{H}_{-\vec{k}}\vec{H}_{\vec{k}}\biggl)
\end{equation}

Thus, the Hamiltonian radiation $\hat{H}_R $ is determined for the Bose gas consisting of the Bose-particles with wave numbers $ k\leq k_0$: 
\begin{equation}
\hat{H}_R=\hat{H}_e+ \hat{H}_h
\end{equation}
where $\hat{H}_e$ and $\hat{H}_h$ are presented by formulas (41) and (43).

To evaluate an energy levels of the operators $\hat{H}_R$ in (44)
within diagonal form, we again apply new linear transformation:

\begin{equation}
\vec {E}_{\vec{k}}=\vec {H}_{\vec{k}}=\frac{\vec {h}_{\vec{k}} + 
M_{\vec{k}}\vec {h}^{+}_{-\vec{k}}} {\sqrt{1-M^2_{\vec{k}}}}
\end{equation} 

where $M_{\vec{k}}$ is the real symmetrical functions  
of  a wave vector $\vec{k}$.

which transforms a form of operator Hamiltonian $\hat{H}_R$ by following way:  
\begin{equation}
\hat{H}_R=
2\sum_{ k\leq k_0}\eta_{\vec{k}}\vec {h}^{+}_{\vec{k}} 
\vec {h}_{\vec{k}}
\end{equation}

Hence, we infer that the Bose-operators 
$\vec {h}^{+}_{\vec{k}}$ and 
$\vec {h}_{\vec{k}}$ are, respectively, 
the vector operators "creation" and "annihilation" of free photons with energy 
 
\begin{equation}
\eta_{\vec{k}}=
\sqrt{\biggl (\frac{\hbar^2k^2}{2m}+\frac{\beta^2}{4\pi }\biggl )^2-\biggl (\frac{\hbar^2k^2}{2m}-\frac{\beta^2}{4\pi }\biggl)^2}=\frac{\hbar k\beta }{\sqrt{2m\pi}}=\hbar k c
\end{equation}

where $\vec {h}^{+}_{\vec{k}}\vec {h}_{\vec{k}}$ is the scalar operator of the number photons occupying the wave vector $\vec{k}$; $c$ is the velocity of photon which defines $\beta =c\sqrt{2m\pi}$ because $c=\frac{ \beta }{\sqrt{2m\pi}}$ in (47). In this respect, the maximal wave number equals to $k_0=\frac{\beta }{\hbar}\sqrt{\frac{m}{2\pi}}=\frac{mc}{\hbar}$.   

Thus, the quantized Maxwell equations have following forms:

\begin{equation} 
\vec {E}=- \frac{\hbar \sqrt{2\pi}}{c\sqrt{m}}\cdot
\frac{d {\vec{H}_0}}{d t}+ c\sqrt{2m\pi}\cdot \vec {E}_0
\end{equation} 
and
\begin{equation} 
\vec {H}=\frac{\hbar \sqrt{2\pi}}{c\sqrt{m}}\cdot
\frac{d {\vec{E}_0}}{d t} + c\sqrt{2m\pi} \vec{H}_0
\end{equation}

\vspace{5mm}
{\bf V. NEW FUNDAMENTAL CONSTANT IN THE NATURE.}
\vspace{5mm}

Our investigation showed that the boson of electromagnetic field has a certainly finite mass $m$. To find the later we states that the source of the photon modes are been the chemical elements which may consider as an ion+electron system which are like to the Hydrogen atom. Due to changing of a electron of its energetic level, by going from high level to low one, leads to an appearance of a photon with energy is determined by a distance between energetic states. The ionization energy of the Hydrogen atom $E_I=\frac{m_e e^4}{2\hbar^2}$ (where $m_e$ and $e$ are the mass and charge of electron) defines the energy of the radiated photon by maximal wave-number $k_0$. Therefore, we may suggest that $\frac{m_e e^4}{2\hbar^2}=\hbar k_0 c$ where $ k_0=\frac{mc}{\hbar}$. This fact discovers a new fundamental constant, which represents as a mass of the light boson:

$$
m=\frac{m_e e^4}{2\hbar^2 c^2}=2.4 \cdot 10^{-35} kg
$$    

In conclusion, we can note that four fundamental particles exist in the nature: 1. the electron with mass $m_e=9\cdot 10^{-31} kg $; 2. the proton with mass $m_p=1.6\cdot 10^{-27} kg $; 3. the neutron with $m_n=1.6\cdot 10^{-27} kg $;  4. the light boson  with mass  $ m=2.4 \cdot 10^{-35} kg $.

Now, we present the form of the interaction energy $\hat{U}_{\vec{r}}$ between two light bosons in the coordinate space in (41), at 
$$
\hat{U}_{\vec{k}}=-\frac{\hbar^2k^2 V}{2m }+
\frac{m c^2 V}{2} >0
$$  
Our calculation shows that

\begin{eqnarray}
\hat{U}_{\vec{r}}&=&\frac{1}{V}\sum_{ k\leq k_0} \hat{U}_{\vec{k}}\cdot e^{i\vec{k}\vec{r}}=4\pi \int^{k_0}_{0} k^2\hat{U}_{\vec{k}} \frac{sin (kr)}{kr}dk=
\nonumber\\
&=&\frac{2\pi V m c^2}{r^3}\biggl[sin \biggl (\frac{mc r}{\hbar}\biggl)+ \frac{mc r}{\hbar} - \frac{mc r}{\hbar} cos \biggl (\frac{mc r}{\hbar}\biggl)\biggl]- 
\nonumber\\
&-&\frac{2\pi V \hbar^2}{m r^5}\biggl [\biggl(\frac{3m^2 c^2 r^2}{\hbar^2}-6\biggl) \cdot sin \biggl (\frac{mc r}{\hbar}\biggl)+ \frac{ m^3 c^3 r^3}{\hbar^3} - \frac{6mc r}{\hbar} -
\nonumber\\
&-&\biggl(\frac{ m^3 c^3 r^3}{\hbar^3} - \frac{6mc r}{\hbar} \biggl)\cdot cos \biggl(\frac{mc r}{\hbar}\biggl) \biggl]
\end{eqnarray}

The existence of a boundary wave number $k_0=\frac{mc}{\hbar}$ for electromagnetic field is connected with the characteristic length of the interaction $\hat{U}_{\vec{r}}$ between two light bosons in the coordinate space  that is a minimal distance $d=\frac{\hbar}{mc}=2.6\cdot 10^{-8} m$ between two neighboring light bosons into the electromagnetic field. We may state herein that the total number of light bosons in volume $V$ is determined as 
$$
\frac{V}{N}=\frac{4\pi d^3}{3}
$$

from which $\frac{N}{V}=1.4\cdot 10^{22} m^{-3}$.

\vspace{5mm}
{\bf V1. SUPERFLUID COMPONENT OF ELECTROMAGNETIC FIELD.}
\vspace{5mm}

The connection between the ideal Bose gas and superfluidity in helium was first made by London [4] in 1938. He postulated that the ideal Bose gas undergoes a phase transition at sufficiently low temperatures to a condition in which the zero-momentum quantum state is occupied by a finite fraction of the light bosons. This momentum-condensed phase was postulated by London to represent the superfluid component of liquid $^4$He. Hence, we postulate that in the Bose gas consisting of $N$ neutral light bosons with spin one and mass $m$, a finite fraction of the light bosons is occupying the zero-momentum quantum state, which determines a superfluid component for the electromagnetic field. 

In this respect, the Hamiltonian radiation in (44) is proposed as:

\begin{eqnarray} 
\hat{H}_R&=&
2\sum_{ k\leq k_0}\biggl(\frac{\hbar^2
k^2}{2m }+
\frac{mc^2}{2}\biggl )
\vec{H}^{+}_{\vec{k}}\vec{H}_{\vec{k}}- 
\nonumber\\
&-&2\sum_{ k\leq k_0}\biggl (\frac{\hbar^2
k^2}{4m }-\frac{mc^2}{4}\biggl )
\biggl (\vec{H}^{+}_{\vec{k}}
\vec{H}^{+}_{-\vec{k}}+   
\vec{H}_{-\vec{k}}\vec{H}_{\vec{k}}\biggl)
\end{eqnarray}

On other hand, the law conservation for total number of light bosons is determined as

\begin{equation} 
N_0+\sum_{0< k\leq k_0}
\vec{E}^{+}_{\vec{k}}\vec{E}_{\vec{k}}+\sum_{0< k\leq k_0}
\vec{H}^{+}_{\vec{k}}\vec{H}_{\vec{k}}=N
\end{equation}

where   $\vec {E}^{+}_{\vec{k}}\vec {E}_{\vec{k}}$ and $\vec {H}^{+}_{\vec{k}}\vec {H}_{\vec{k}}$ are, respectively, the scalar operator of the number of the light bosons of electric and magnetic fields with spin one occupying the wave vector $\vec{k}$ which satisfy to the condition$\vec {E}^{+}_{\vec{k}}\vec {E}_{\vec{k}}= vec {H}^{+}_{\vec{k}}\vec {H}_{\vec{k}}$; $N_0=\vec{E}^{+}_0\vec{E}_0+\vec{H}^{+}_0\vec{H}_0$ is the scalar operator of number of the light bosons of electric and magnetic fields in the condensate.

Now let us derive an equation for the density of condensed light bosons.
In statistical equilibrium, the equation for the density of condensate light bosons is represented as
\begin{equation}
N_{0, T} +2\sum_{0< k\leq k_0}
\overline{\hat{H}^{+}_{\vec{k}}\hat{H}_{\vec{k}}}  = N
\end{equation}

$N_0= N_{0, T}$ is the average number of condensed light bosons at the temperature $T$; $\overline{\hat{H}^{+}_{\vec{k}}\hat{H}_{\vec{k}}}$
is the average number of light bosons of magnetic field which equals to the average number of light bosons of electric field with the wave vektor $\vec{k}$ at a
temperature $T$.

To get the form $\overline{\hat{H}^{+}_{\vec{k}}\hat{H}_{\vec{k}}}$ we use of new linear transformation presented in (45). Then,
\begin{equation}
\overline{\hat{H}^{+}_{\vec{k}}\hat{H}_{\vec{k}}}=
\frac{1+M^2_{\vec{p}}}{1-M^2_{\vec{p}}}\overline{\hat{h}^{+}_{\vec{k}}\hat{h}_{\vec{k}}}+
\frac{M_{\vec{k}}}{1-M^2_{\vec{k}}}\biggl(\overline{\hat{h}^{+}_{\vec{k}}\hat{h}^{+}_{-\vec{k}}} +
\overline{\hat{h}_{\vec{h}}\hat{h}_{-\vec{k}}}\biggl) + \frac{M^2_{\vec{k}}}{1-M^2_{\vec{k}}}
\end{equation}
where $\overline{\hat{h}^{+}_{\vec{k}}\hat{h}_{\vec{k}}}$
is the average number of photon modes with the wave vector $\vec{k}$ at a
temperature $T$:

\begin{equation}
\overline{\hat{h}^{+}_{\vec{k}}\hat{h}_{\vec{k}}}=\frac{1}{e^{\frac{\eta_{\vec{k}}}{kT}}-1}
\end{equation}

On other hand, by the theorem of Bloch-De-Dominisis:

\begin{equation}
\overline{\hat{h}^{+}_{\vec{k}}\hat{h}^{+}_{-\vec{k}}}=
\overline{\hat{h}_{\vec{k}}\hat{h}_{-\vec{k}}}=0
\end{equation}
Consequently, the equation for density light bosons in the condensate (51), by using of  (54) and (56), takes a following form:

\begin{equation}
\frac{N_{0, T} }{V}=\frac{N }{V}-
\frac{2}{V}\sum_{0< k\leq k_0}\frac{M^2_{\vec{k}}}{1-M^2_{\vec{k}}}-
\frac{2}{V}\sum_{0< k\leq k_0}\frac{1+M^2_{\vec{k}}}{1-M^2_{\vec{k}}}
\overline{\hat{h}^{+}_{\vec{k}}\hat{h}_{\vec{k}}}
\end{equation}

where the real symmetrical functions $M_{\vec{k}}$ from a wave vector $\vec{k}$ equals to:

\begin{equation}
M^2_{\vec{k}}=\frac{\frac{\hbar^2
k^2}{2m }+
\frac{mc^2}{2}-\hbar k c}{\frac{\hbar^2
k^2}{2m }+
\frac{mc^2}{2}+\hbar k c }
\end{equation}

In conclusion, we note that the electromagnetic field is considered as superfluid liquid, at absolute zero $T=0$, because the average number of photons (55) is zero $\overline{\hat{h}^{+}_{\vec{k}}\hat{h}_{\vec{k}}}=0$, and

\begin{equation}
\frac{N_{0, 0} }{V}=\frac{N }{V}-\frac{1}{\pi^2}\int^{k_0}_{0}k^2\frac{M^2_{\vec{k}}}{1-M^2_{\vec{k}}}dk=\frac{N }{V}-\frac{ m^3 c^3}{4\pi^2\hbar^3}B(2,3)
 \end{equation}
where $ B(2,3)$ is the beta function
$$
B(2,3)=\int^{1}_{0}x \biggl(1-x\biggl )^2dx=\frac{1}{10}
$$
Thus, the condensed fraction $\frac{N_{0, 0} }{N}$, at absolute zero, equals to
$\frac{N_{0, 0}}{N}=0.9$, at application of a values as $\frac{N}{V}=1.4\cdot 10^{22} m^{-3}$;  $k_0=\frac{mc}{\hbar}$; $ m=2.4 \cdot 10^{-35} kg $. 

The transition temperature $ T=T_{\lambda}$ for electromagnetic field is determined within transformation of a superfluid phase of the light boson gas by the normal phase, when  the density light bosons in the condensate satisfies to the condition $\frac{N_{0, T_{\lambda}} }{V}=0$. This fact presents a following equation as result of (57) by using of (59):

\begin{equation}
\frac{N_{0, 0} }{N}=
\frac{ k^4 T^4_{\lambda}V}{2\pi^2\hbar^3 m c^5 N}\int^{\alpha}_{0}\frac{x^3 dx}{e^{x}-1}+\frac{k^2 m T^2_{\lambda}V}{2\pi^2\hbar^3 c N}\int^{\alpha}_{0}\frac{x dx}{e^{x}-1}
\end{equation}

where $\alpha =\frac{mc^2}{k T_{\lambda}}$.

To solve an equation (58) in regarding to  $ T_{\lambda}$ we consider the certainly case when $\alpha =\frac{mc^2}{k T_{\lambda}}\gg 1$. In this respect, 
$$
\int^{\alpha}_{0}\frac{x^3 dx}{e^{x}-1}=\int^{\infty }_{0}\frac{x^3 dx}{e^{x}-1}=\frac{\pi^4}{15}
$$
and
$$
\int^{\alpha}_{0}\frac{x dx}{e^{x}-1}=\int^{\infty }_{0}\frac{x dx}{e^{x}-1}=\frac{\pi^2}{6}
$$
Our calculation shows that the transition temperature for electromagnetic field is $ T_{\lambda}\approx  10^4 K$ which satisfies to the condition $\alpha =\frac{mc^2}{k T_{\lambda}}\gg 1$.

\newpage 
\begin{center} 
{\bf References} 
\end{center} 
 
\begin{enumerate} 
\item 
P.A.M..~Dirac~, "The Principles of Quantum Mechanics", ~Oxford at the  
Clarendon  press  (1958), "Lectures on Quantum Mechanics", ~Yeshiva University New York~  (1964)  
\item
N.N.~Bogoliubov~, Jour. of Phys.(USSR), ~{\bf 11},~23~(1947)
\item 
A.~Korn~, M. ~Korn ~, "Mathematical Handbook", ~McGraw –Hill Book company  
(1968)
\item
F.~London~, Nature, ~{\bf 141},~643~(1938)

\end{enumerate} 
\end{document}